\documentclass{iau}
\usepackage{graphicx}
\title[Simulations with MPC] 
{Multiparticle collision simulations of dense stellar systems and plasmas}
\author[P. Di Cintio et al.]   
{P. Di Cintio$^{1,2,3}$, M. Pasquato$^{4,5}$, L. Barbieri$^{2,3,6}$, H. Bufferand$^7$,\\
L. Casetti$^{2,3,6}$, G. Ciraolo$^7$, U. N. di Carlo$^8$, P. Ghendrih$^7$,\\ 
J. P. Gunn$^7$, S. Gupta$^9$, H. Kim$^{10}$, S. Lepri$^{1,3}$, R. Livi$^{2,3,1}$,\\
A. Simon-Petit$^{2,3}$, A. A. Trani$^{11,12}$, S.-J. Yoon$^{10}$}
\affiliation{$^1$Consiglio Nazionale delle Ricerche, Istituto dei Sistemi Complessi via Madonna del piano 10, I-50019 Sesto Fiorentino, Italy \\ email: {\tt pierfrancesco.dicintio@cnr.it} \\[\affilskip]
$^2$Dept. of Physics \& Astronomy, University of Florence, via G. Sansone 1, I-50019 Sesto Fiorentino (FI), Italy\\[\affilskip]
$^3$INFN, Sezione di Firenze, via G. Sansone 1, I-50019 Sesto Fiorentino (FI), Italy\\[\affilskip]
$^4$Dept. of Physics \& Astronomy Galileo Galilei, University of Padova, Vicolo dell'Osservatorio 3, I-35122, Padova, Italy \\[\affilskip]
$^5$Département de Physique, Université de Montréal, Montreal, Quebec H3T 1J4, Canada\\[\affilskip]
$^6$INAF, Osservatorio Astrofisico di Arcetri, Largo Enrico Fermi, 5, I-50125 Firenze, Italy\\[\affilskip]
$^7$ IRFM, CEA, F-13108 St Paul Lez Durance, France\\[\affilskip]
$^8$McWilliams Center for Cosmology and Department of Physics, Carnegie Mellon University, Pittsburgh, PA 15213, USA\\[\affilskip]
$^9$ Tata Institute of Fundamental Research, Homi Bhabha Road, Colaba Mumbai 400005, India\\[\affilskip]
$^{10}$Department of Astronomy \& Center for Galaxy Evolution Research, Yonsei University,\\ Seoul 120-749, Republic of Korea\\[\affilskip]
$^{11}$Department of Earth Science and Astronomy, College of Arts and Sciences, The University of Tokyo, 3-8-1 Komaba, Meguro-ku, Tokyo 153-8902, Japan\\[\affilskip]
$^{12}$Okinawa Institute of Science and Technology, 1919-1 Tancha, Onna-son, Okinawa 904-0495, Japan}
\pubyear{2022}
\volume{362}  
\pagerange{yyy--zzz}
\jname{Predictive Power of Computational Astrophysics as a Discovery Tool}
\editors{D. Wiebe, T. Hanawa, C. Boily \& J. Stone, eds.}
\begin{document}
\maketitle
\begin{abstract}
We summarize a series of numerical experiments of collisional dynamics
in dense stellar systems such as globular clusters (GCs) and in weakly collisional plasmas using a novel simulation technique, the so-called Multi-particle collision (MPC) method,
alternative to Fokker-Planck and Monte Carlo approaches.
MPC is related to particle-mesh approaches for the computation of self consistent long-range fields, ensuring that simulation time scales with $N\log N$ in the number of particles, as opposed to $N^2$
for direct $N$-body. The collisional relaxation effects are modelled
by computing particle interactions based on a collision operator
approach that ensures rigorous conservation of energy and momenta and depends only on particles velocities and cell-based integrated quantities.
\end{abstract}
\firstsection 
\section{Introduction}
The evolution of self-gravitating systems and plasmas is mainly dominated by collective mean-field processes due to their large number of particles $N$ and the long-range nature of the $1/r^2$ gravitational or Coulombian force. The typical dynamical scale $t_{\rm dyn}$ of such systems is of the order of $1/\sqrt{G\bar\rho}$, where $G$ is the gravitational constant and $\bar\rho$ is some mean mass density, for the self-gravitating systems, and proportional to the inverse of the plasma frequency $\omega_P=\sqrt{4\pi n_e e^2/m_e}$, where $n_e$, $e$ and $m_e$ are the number density, charge and mass of the electrons, for plasmas.\\  
\indent Collisional processes typically contribute to the dynamics on a time scale that is a function\footnote{Note that $N$ corresponds to the total number of particles of the systems in the gravitational case, while it is the average number of particles, ions or electrons, withing a Debye length for plasmas.} of $N$ as $t_{c}\approx t_{\rm dyn}\ln{N}/N$, that in the case of gravitating systems may exceed several times the age of the Universe (see \cite{bt08}), while in plasma physics is always regulated by temperature and mean free path. For these reasons, the numerical modelling of systems governed by $1/r^2$ forces is usually carried by means of {\it collisionless} approaches such as the widely uses particle-mesh or particle-in-cell (PIC) schemes (see e.g. \cite{he88,dr11}). However, there are several  examples of of systems that are (at least partially) dynamically regulated by collisions. In particular, dense stellar systems such as globular clusters or galactic cores can be largely in collisional regimes, while their modelling in terms of ``honest" direct $N-$body simulations with a one-to-one correspondence between stars and simulation particles remains still challenging due to the (relatively) large size $N$ of the order of $10^6$. In plasma physics, collisional systems can be found in trapped one component ion or electron plasmas (see \cite{dub99}) or ultracold neutral plasmas (see \cite{pohl05}), while a transition between collisional and collisionless plasma regimes is expected in the scrape-off layer of tokamaks (see \cite{funda05}).\\
\indent Numerical simulations of collisional stellar systems usually employ the Monte Carlo technique or hybrid particle-mesh direct $N-$body schemes. Vice versa, PIC plasma codes based on the solution of the Vlasov-Maxwell equations on some reduced geometry rely on analytic or semi-analytic methods to reconstruct the collision integral in the kinetic picture.\\
\indent In a series of papers, we have implemented a novel approach to the simulation of both gravitational and Coulomb collisional systems based on the so-called multiparticle collision operator (hereafter MPC), originally developed by \cite{kapral99} in the context of mesoscopic fluid dynamics (see also \cite{gomper09} and references therein for an extensive review). Here we give a brief introduction of the method and highlight the major results obtained so far.
\section{Overview of the method}
The MPC scheme in a three-dimensional simulation domain containing $N$ particles partitioned in $N_c$ cells, amounts to a cell-dependent rotation of an angle $\alpha_i$ of the particle's relative velocity vectors $\delta\mathbf{v}_j=\mathbf{v}_j-\mathbf{u}_{\rm{com},i}$ in the centre of mass frame of each cell $i$, so that 
\begin{equation}\label{rotation}
\mathbf{v}_{j}^\prime=\mathbf{u}_{\rm{com},i}+\delta\mathbf{v}_{j,\perp}{\rm cos}(\alpha_i)+(\delta\mathbf{v}_{j,\perp}\times\mathbf{R}_i){\rm sin}(\alpha_i)+\delta\mathbf{v}_{j,\parallel}.
\end{equation}  
In the formulae above $\mathbf{R}_i$ is a random axis chosen for the given cell,  $\delta\mathbf{v}_{j,\perp}$ and $\delta\mathbf{v}_{j,\parallel}$ are the relative velocity components perpendicular and parallel to $\mathbf{R}_i$, respectively and
\begin{equation}
  \mathbf{u}_{\rm{com},i}=\frac{1}{m_{{\rm tot},i}}\sum_{j=1}^{n_i}m_j\mathbf{v}_j;\quad m_{{\rm tot},i}=\sum_{j=1}^{n_i}m_j.
\end{equation}
For each cell, if the rotation angle $\alpha_i$ is chosen randomly by sampling a uniform distribution, after the vectors $\delta\mathbf{v}_j$ are rotated back around $\mathbf{R}_i$ by $-\alpha_i$, the linear momentum and the kinetic energy are conserved {\it exactly} in each cell, from which follows the conservation of the {\it total} parent quantities (see e.g. \cite{dicintio17}), at variance with other previously implemented collision schemes, such as for example the \cite{nambu83} method that imposes the conservation of the kinetic energy only, while breaking the conservation of linear momentum (that is preserved only in time average).\\
\indent By introducing an additional constraint on the rotation angles $\alpha_i$, one can add the  conservation of a component of the total angular momentum of the cell $\mathbf{L}_i$ (see e.g. \cite{noguchi08}). In this case  $\alpha_i$ is given by
\begin{equation}\label{sincos}
{\rm sin}(\alpha_i)=-\frac{2a_ib_i}{a_i^2+b_i^2};\quad  {\rm cos}(\alpha_i)=\frac{a_i^2-b_i^2}{a_i^2+b_i^2},
\end{equation}
where
\begin{equation}\label{ab}
a_i=\sum_{j=1}^{N_i}\left[\mathbf{r}_j\times(\mathbf{v}_j-\mathbf{u}_i)\right]|_z;\quad b_i=\sum_{j=1}^{N_i}\mathbf{r}_j\cdot(\mathbf{v}_j-\mathbf{u}_i).
\end{equation}
In the expression above $\mathbf{r}_j$ are the particles position vectors, and with $[\mathbf{x}]|_z$ we assume that we are considering (without loss of generality) the component of the vector $\mathbf{x}$ parallel to the $z$ axis of the simulation's coordinate system, that implies that in our case the $z$ component of the cell angular momentum is conserved.\\
\indent The exact conservation of the total angular momentum inside the cell can be enforced by choosing $\mathbf{R}_i$ parallel to the direction of the cell's angular momentum vector $\mathbf{L}_i$ and using the definition of $a_i$ accordingly.\\
\indent In our implementation of the MPC method, the collision step is always conditioned to a cell-dependent probability accounting for "how much the system is collisional" locally. The latter is defined as 
\begin{equation}\label{cumulative}
p_i={\rm Erf}\left(\beta\Delta t \nu_c\right),
\end{equation}
where $\Delta t$ is the simulation timestep, $\nu_c$ is the collision frequency, $\beta$ is a dimensionless constant of the order of twice the number of the simulation cells, and ${\rm Erf}(x)$ is the standard error function. In Equation (\ref{cumulative}) the collision frequency is given by
\begin{equation}
\nu_c=\frac{8\pi G^2\bar{m}^2_i\bar n\log\Lambda}{\sigma^3_i}
\end{equation}
for the gravitational case, where $\bar{n}$ the mean stellar number density, $\bar{m}_i$ and $\sigma_i$ the average particle mass and the velocity dispersion in the cell, respectively, and by
\begin{equation}
\nu_c=\frac{4\pi e^4n_e\log\Lambda}{m_e^{1/2}T^{3/2}}
\end{equation}
for the plasma case, where $T$ is the electron temperature. In both cases the Coulomb logarithm $\log\Lambda$ of the maximum to minimum impact parameter is usually taken of order 10 for the systems of interest. Otherwise, it can also be evaluated cell-dependently.\\ 
\indent Once Equation (\ref{cumulative}) is evaluated in each cell, a random number $p_{*i}$ is sampled from a uniform distribution in the interval $[0,1]$ and the multi-particle collision is applied for all cells for which $p_{*i}\leq p_i$.\\
\indent Between two applications of the MPC operator, the particles are propagated under the effect of the self-consistent gravitational or electromagnetic fields, obtained by solving at on a grid the Poisson or Maxwell equations with standard finite elements schemes. Form the point of view of the computational cost, for a given initial condition and fixing the order of the particle integrator and time step, the MPC simulations are on average a factor 5 faster than direct $N-$body calculation in the range of $N$ between $10^3$ to $10^6$. Adding the conservation of angular momentum and including the binaries (or processes of electron capture or collisional ionization in the plasma codes) typically reduces the computationl gain of a factor two with respect to other codes.

\section{Results and discussion}
\begin{figure}
  \includegraphics[width=0.9\textwidth]{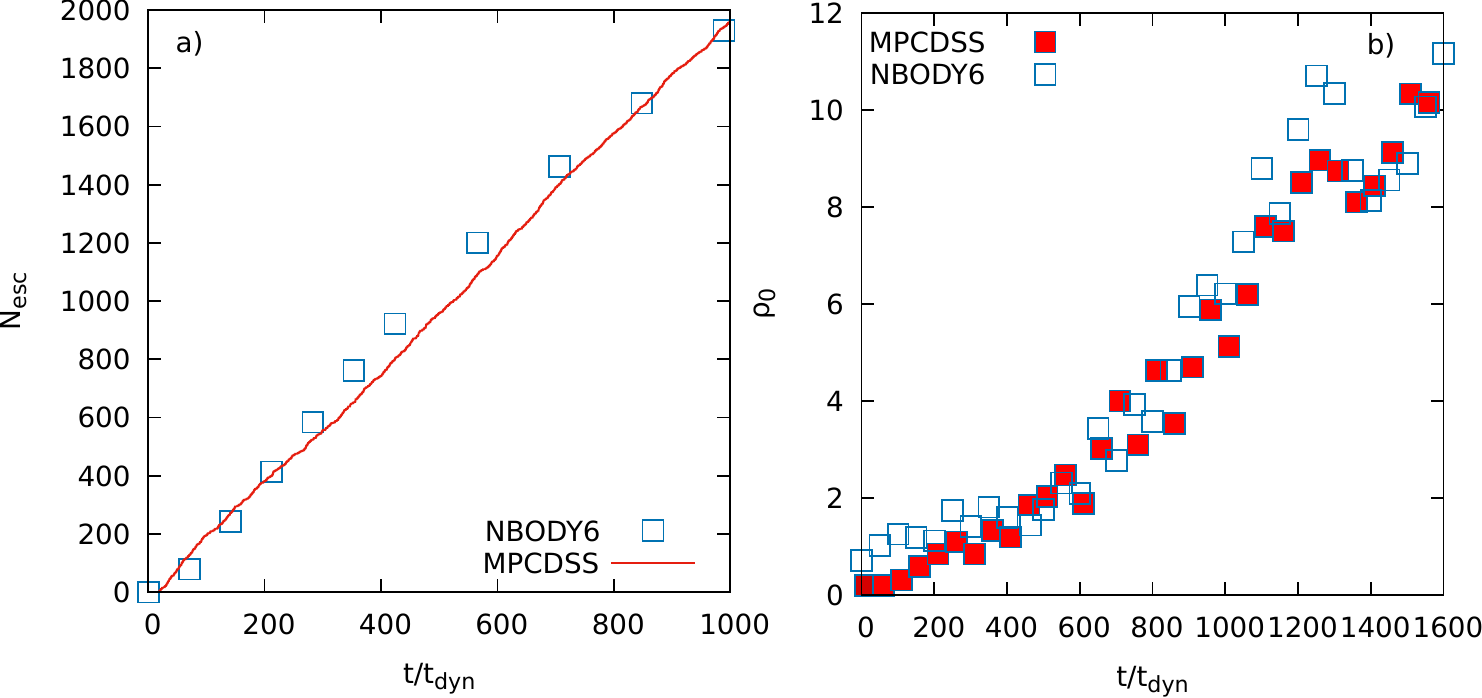}
\caption{a) Number of escapers as function of time for a cluster with  $N=32000$ particles propagated with {\sc mpcdss} (solid line) and {\sc nbody6} (squares). b) Evolution of the central density for a core collapsing cluster evolved with {\sc mpcdss}  (filled symbols) and {\sc nbody6} (empty symbols).}
\label{figmpcdss}       
\end{figure}
\subsection{Evolution of Globular clusters}
\cite{dicintio21} and \cite{dicintio22a} performed a wide range of hybrid MPC-Particle-mesh simulations of globular clusters with the {\sc mpcdss} code, investigating the effect of a mass spectrum on the dynamics of core collapse, with particle numbers up to $N=10^6$, different radial anisotropy profiles while adopting the widely used Plummer density distribution. Low resolution simulations with $N=32000$ have been compared to their counterparts performed with the direct $N-$body code {\sc nbody6} (see \cite{aarseth}) using the exact same initial conditions. Overall we observed a good agreement between the two approaches in the low $N$ limit in the evolution of indicators such as the fraction of escapers (i.e. particles with {\it positive} total energy being outside a given cut-off radius) shown in panel a) of Fig. \ref{figmpcdss} or the mean density $\rho_0$ evaluated within a fixed radius of the order one tenth of the initial scale radius, see panel b) same figure. The slight discrepancy between the evolution of said central density observed a late stages of the collapse can be ascribed to the fact that the MPC and the $N-$body simulations have radically different force-evaluation schemes, and, due to the intrinsically chaotic nature of the $N-$body problem, particle trajectories can differ sensibly even starting with analogous initial conditions. This reflects in the fact that denser regions where particle encounters happen more frequently could show rather complex fluctuations.\\ 
\indent In models with a mass spectrum we confirm the theoretical self-similar contraction picture but with a dependence on the slope of the mass function. Moreover, the time of core collapse shows a non-monotonic dependence on the slope that holds also for the depth of core collapse and for the dynamical friction timescale of heavy particles. Cluster density profiles at core collapse show a broken power law structure, suggesting that central cusps are a genuine feature of collapsed cores.\\
\indent In addition we also investigated the dynamics of a central intermediate mass black hole (IMBH) finding that the latter, independently on the structure of the mass spectrum and the anisotropy profiles accelerates the core collapse while making it shallower. In general we also observe that the presence of a mass spectrum results in a way large wander radius of the IMBH for fixed total stellar mass and different masses of the IMBH itself.\\
\indent The results on the structure and evolution of velocity dispersion and anisotropy profiles of cluster with central IMBHs are to be published elsewhere (\cite{dicintio22b}).
\begin{figure}
  \includegraphics[width=0.85\textwidth]{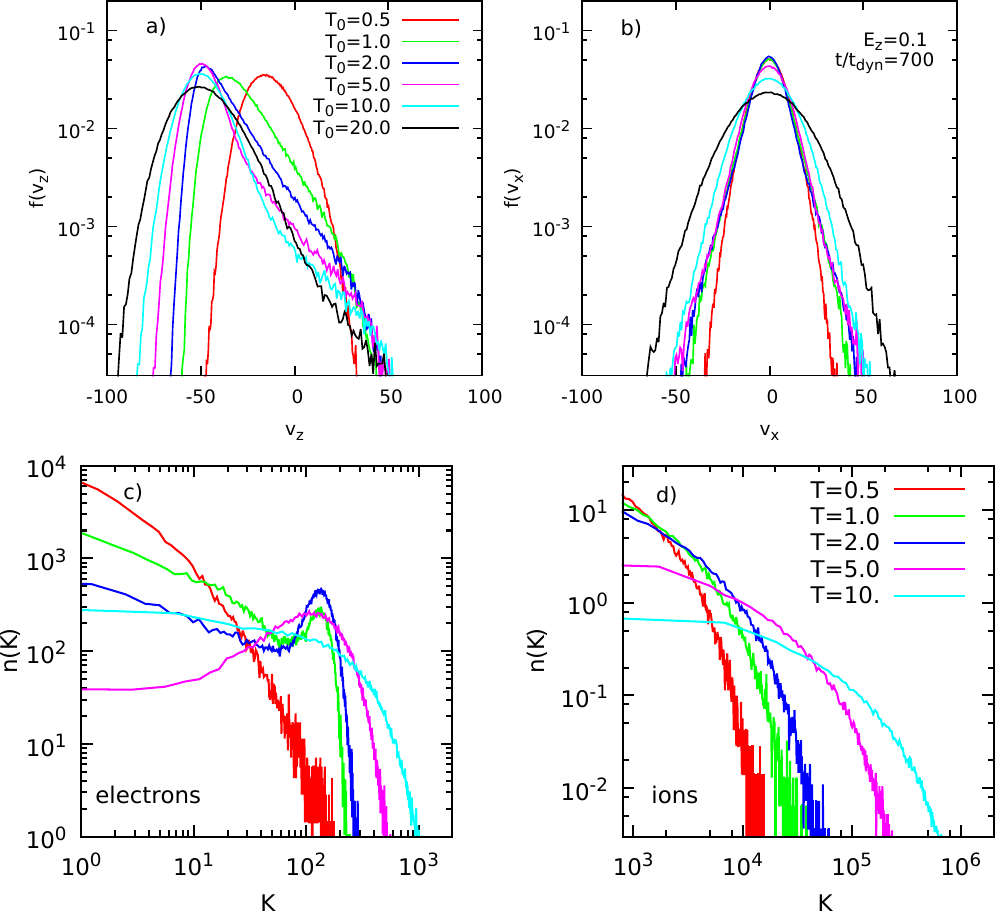}
\caption{a) Electron velocity distribution along the direction of the electric field. b) Electron velocity distribution in the transverse direction. c) Electron kinetic energy distribution. d) Ion kinetic energy distribution. All distributions are evaluated at $t\sim 700/\omega_P$ for different values of $T_0$ (indicated in figure) and $E_z=0.1$ in computer units.}
\label{figtropico}       
\end{figure}
\subsection{Weakly collisional plasmas}
The most part of the MPC simulations performed with the {\sc tropic$^3$o} code (see \cite{dicintio15}, \cite{dicintio17}) where devoted to the investigation of the transition between different regimes of anomalous energy transport in low dimensional (1D and 2D, see \cite{lepri19}) toy set-ups aiming at shedding some light on the heat flux profile structure along magnetic field lines crossing strong temperature gradients in weakly collisional plasmas, relevant for magnetic fusion devices. \cite{ciraolo18} found a surprisingly good agreement between the electron temperature profiles between a hot source and a cold interface obtained by MPC simulations with the corresponding profiles evaluated by 1D fluid codes with a semi-analytical collisional closure of the fluid equations and a non-local definition of the heat flux accounting for suprathermal particles.\\
\indent \cite{lepri21} further investigated the nonequilibrium steady states of a 1D models of finite length L in contact with two thermal-wall heat reservoirs, finding a clear crossover from a kinetic transport regime to an anomalous (hydrodynamic) one over a characteristic scale proportional to the cube of the collision time among particles. In addition, test simulations of  models with thermal walls injecting particles with given non thermal velocity. showed that for fast and relatively cold particles, smaller systems never establish local equilibrium keeping non-Maxwellian velocity distributions.\\
\indent \cite{dicintio18} used MPC simulations to study the non thermal profiles of the weakly ionized media in filamentary structures in molecular clouds finding that strong collisions in dense regions enforce the production of high velocity tails in the particle's distribution, able to climb the gravitational potential well of the filament, thus forming hot and diffuse external envelopes without additional heat sources from the intergalactic media or star formation feedback. Density and temperature profiles obtained in these numerical experiments qualitatively mach the observed transverse profiles of both gas density and temperature.\\
\indent The possibility to incorporate an energy and momentum preserving particle-based collisional operator in plasma codes opens to the possibility to study problems involving the presence of different species or complex non thermal processes without making strong assumptions on a (possibly unknown) phase-space distribution function. In particular, we aim at studying the mechanism of electron run-away in presence of net electrostatic fields and/or violation of the plasmas quasineutrality. We performed some preliminary test simulations in a 3D periodic geometry aiming at observing the formation of a suprathermal tail in the velocity distribution when applying a constant electric field $\mathbf{E}$ along one of the three spatial coordinates, for different values of the initial electron temperature $T_0$ at fixed density $n_e$ and assuming that electrons and ions are initially at equilibrium. In Fig. \ref{figtropico} we show the electron velocity distributions along the direction of $\mathbf{E}$ ($z$, without loss of generality) and one of the transverse directions $f(v_z)$ and $f(v_x)$, panels a) and b). It appears clearly that for fixed density and therefore fixed mean free path, systems with lower initial electron temperature are less and less  prone to develop non thermal tails along the direction of the fixed external field as particles accelerated by such field suffer a stronger dynamical friction effect due to the lower velocity dispersion, as theorized by \cite{dreicer59}. Further increasing $T_0$ produces more and more deformed $f(v_z)$ while the velocity distribution in the transverse directions $f(v_x)$ regains a thermal structure when $T_0$ is larger than some critical value for which the hotter systems are basically collisionless and the different degrees of freedom are decoupled. This is also evident from the full differential kinetic energy distribution $n(K)$ (see panel c, same figure), where the ''mono-chromatic" peak at $K\approx1.5\times 10^2$ in computer units appears only for systems starting with somewhat intermediate values of $T_0$. Being heavier (in this case with a mass ratio of $2\times 10^3$), and thus having lower mobility, ions do not bare any peculiar structure in their kinetic energy distributions even at late times of the order of thousand of plasma oscillations, such that their $n(K)$ corresponding to different $T_0$ can be collapsed onto one another (panel d).


\begin{thebibliography}{}



\bibitem[Binney \& Tremaine (2008)]{bt08} 
{Binney, J.; Tremaine, S.} 2008 
Galactic dynamics (2nd edition, Princeton University Press)


\bibitem[Ciraolo \etal\ (2018)]{ciraolo18}
{Ciraolo, G., Bufferand, H., Di Cintio, P., Ghendrih, P., Lepri, S., Livi, R., Marandet, Y., Serre, E., Tamain, P. \& Valentinuzzi, M.} 2018, 
\textit{Contributions to Plasma Physics}, 58, 457



\bibitem[Dehnen \& Read (2011)]{dr11} 
{Dehnen, W. \& Read, J.I.} 2011, \textit{EPJ
Plus} 126, 55


\bibitem[Di Cintio \etal\ (2015)]{dicintio15}
{Di Cintio, P., Livi, R., Bufferand, H., Ciraolo, G. Lepri, S. \& Straka M.J.} 2015,
\textit{Phys.Rev.E}, 92, 62108 

\bibitem[Di Cintio \etal\ (2017)]{dicintio17}
{Di Cintio, P., Livi, R., Lepri, S. \& Ciraolo, G.} 2017, 
\textit{Phys.Rev.E}, 95, 43203

\bibitem[Di Cintio \etal\  (2018)]{dicintio18}
{Di Cintio, P., Gupta, S., \& Casetti, L.} 2018, 
\textit{MNRAS}, 475, 1137

\bibitem[Di Cintio \etal\  (2021)]{dicintio21}
{Di Cintio, P., Pasquato, M., Kim, H., \& Yoon, S.-J.} 2021, 
\textit{A\&A}, 649, A24

\bibitem[Di Cintio \etal\  (2022a)]{dicintio22a}
{Di Cintio, P., Pasquato, M., Simon-Petit, A., \& Yoon, S.-J.} 2022 (in press), 
\textit{A\&A}

\bibitem[Di Cintio \etal\  (2022b)]{dicintio22b}
{Di Cintio, P., Pasquato, M., Barbieri, L., Trani, A.A., \& di Carlo, U.N.} 2022 (in preparation)

\bibitem[Dreicer (1959)]{dreicer59}{Dreicer, H.} 1959, \textit{Physical Review}, 115, 238


\bibitem[Dubin \& O'neil (1999)]{dub99}{Dubin, D. H. \& O'neil, T. M.} 1999 \textit{Rev. Mod. Phys.}, 71, 87 

\bibitem[Fundamenski (2005)]{funda05}{Fundamenski, W.} 2005 \textit{Plasma Phys. and Contr. Fus.}, 47, 163


\bibitem[Gompper \etal\ (2009)]{gomper09}{G. Gompper, T. Ihle, D. M. Kroll, and R. G. Winkler} 2009, \textit{Multi-Particle Collision Dynamics: A ParticleBased Mesoscale Simulation Approach to the Hydrodynamics of Complex Fluids}, in 	Adv Polym Sci 221, 1


\bibitem[Hockney \& Eastwood (1988)]{he88} 
{Hockney, R. W., Eastwood, J. W.} 1988 
Computer simulations using particles (Taylor \& Francis eds.)

\bibitem[Killian \etal\ (2005)]{pohl05}{Killian, T. C., Pattard, T.,  Pohl, T. \& Rost, J.-M.} 2005 \textit{Physics Reports}, 449,77 


\bibitem[Lepri \etal\ (2019)]{lepri19}{Lepri, S., Bufferand, H., Ciraolo, G., Di Cintio, P., Ghendrih, P. \& Livi, R.} 2019, in Stochastic Dynamics Out of Equilibrium, ed. G. Giacomin, S. Olla, E. Saada, H. Spohn, \& G. Stoltz
(Cham: Springer International Publishing), 364


\bibitem[Lepri \etal\ (2021)]{lepri21}
{Lepri, S., Ciraolo, G., Di Cintio, P., Gunn, J.P. \& Livi, R.} 2021, 
\textit{Phys.Rev.Research}, 3, 13207


\bibitem[Malevanets \&\ Kapral (1999)]{kapral99}
{Malevanets, A. \& Kapral, R.} 1999, 
\textit{J. Chem. Phys}, 110, 8605


\bibitem[Nambu (1983)]{nambu83}
{Nambu, K.} 1983, \textit{J. Phys. Soc. Jap.}, 52, 3382


\bibitem[Nitadori \& Aarseth (2012)]{aarseth}{Nitadori, K., \& Aarseth, S. J.} 2012 \textit{MNRAS} 424, 545


\bibitem[Noguchi \& Gompper (2008)]{noguchi08}{Noguchi, H. \& Gompper, G.} 2008 \textit{Phys. Rev. E} 78, 016706


\end{thebibliography}
\end{document}